\documentclass[conference]{IEEEtran}
\IEEEoverridecommandlockouts
\usepackage{cite}
\usepackage{amsmath,amssymb,amsfonts}
\usepackage{algorithmic}
\usepackage{graphicx}
\usepackage{textcomp}
\usepackage{xcolor}
\usepackage{multirow}
\usepackage{multicol}

\def\BibTeX{{\rm B\kern-.05em{\sc i\kern-.025em b}\kern-.08em
T\kern-.1667em\lower.7ex\hbox{E}\kern-.125emX}}

\usepackage{listings}

\begin{document}

\title{An Experiment of Randomized Hints on an Axiom of Infinite-Valued \L ukasiewicz Logic
}

\author{\IEEEauthorblockN{Ruo Ando}
\IEEEauthorblockA{\textit{National Institute of Informatics} \\
\textit{2-1-2 Hitotsubashi, Chiyoda-ku}\\
\textit{Tokyo, Japan} \\
ruo@nii.ac.jp}
\and
\IEEEauthorblockN{Yoshiyasu Takefuji}
\IEEEauthorblockA{\textit{Musashino University, , Tokyo135-8181, Japan} \\
\textit{3-3-3 Ariake, Koto-Ku} \\
\textit{Tokyo, Japan}\\
}
}

\maketitle

\begin{abstract}
In this paper, we present an experiment of our randomized hints strategy of automated reasoning
for yielding Axiom(5) from Axiom(1)(2)(3)(4) of Infinite-Valued \L ukasiewicz Logic.
In the experiment, we randomly generated a set of hints with size ranging from 30 to 60 for guiding
hyper-resolution based search by the theorem prover OTTER.
We have successfully found the most useful hints list (with 30 clauses) among 150 * 6 hints lists.
Also, we discuss a curious non-linear increase of generated clauses in deducing Axiom(5) by applying our randomized hints strategy.
\end{abstract}


\section{Inifite-valued \L ukasiewicz logic}
Infinite-valued logic is the classic and promising field in automated reasoning.
Recently, Pykacz discussed whether quantum mechanics should be described by many-valued logic \cite{b1}.
Infinite-valued logic can cope with the world without the excluded middle of which principle
Kolmogorov discussed in detail \cite{b2}.
Pablo et al. \cite{Pablo} point out that Kleene's many-valued logic is used to cope with sentences such as Liar or the Truth-Teller.
In the experiment, we use the automated reasoning program OTTER \cite{otter}.
We apply a randomized hints strategy \cite{b5} for directing a hyper-resolution based search.

\subsection{The semantics of infinite value logic}

\L ukasiewicz firstly presented the semantics of the infinite-valued logic (sentential calculus).
His semantics aims to interpret the sentential variables that appeared in
Principia Mathematics by Whitehead and Russell \cite{b4}.
We call the \L ukasiewicz axioms (1)-(5) as $\L_{{\aleph}_{0}}$ corresponding to Harris et al. \cite{b6}.

We introduce a predicate $v$ for assigning rational numbers with the closed unit interval [0,1].
The interval [0,1] is applied to each atomic sentence in $\L_{{\aleph}_{0}}$.
Using $v$, we can cope with the sentences determined recursively, as follows.
\begin{align}
\rightarrow : v(C_{pq}) = p \rightarrow q =
\begin{cases}
{0 \ \text{if $p =>q$}}\\
{q-p \ \text{if $p < q$}}
\end{cases} \\
\& : v(K_{pq}) = max(p,q) \\
\vee : v(A_{pq}) = min(p,q) \\
\neg : 1-p
\end{align}

Here in (1), the predicate $v$ represents a conditional of the $ C_{pq} $, which means p implies q.
In (2), v assigns a conjunction $ K_{pq} $ which yields the maximum of the values of p and q.
In (3), $ A_{pq} $ which is $\vee$ means a disjunction by the minimum of the values of p and q.

Material implecation of \L ukasiewicz logic truth table is as follows:
\begin{table}[h]
\caption{\L ukasiewicz logic truth table.}
\centering
\begin{tabular}{|c|c|ccc|}\hline
\multicolumn{2}{|c|}{ A $ \rightarrow $ B } & \multicolumn{3}{c|}{B}\\
\cline{3-5}
\multicolumn{2}{|c|}{}
& F & U & T \\ \hline
& F & T & T & T \\
A & U & U & T & T \\
& T & F & U & T\\ \hline
\end{tabular}
\end{table}

The \L ukasiewicz logic differs from Kleene logic in its definition of the implication that
``unknown implies unknown'' is true.

\subsection{The Axiomatics of Hilbert-style system $\L_{{\aleph}_{0}}$}

\L ukasiewicz has conjectured that the set of five axiom schemes (A1-A5) and
the single rule of condensed detachment (CD) \cite{kalman}.
With A1-A5 and CD, the semantics of $\L_{{\aleph}_{0}}$ are completed.

\begin{align}
(A1): A \rightarrow (B \rightarrow A) \\
(A2): ((A \rightarrow B) -> (B \rightarrow C)) \rightarrow (A \rightarrow C) \\
(A3): ((A \rightarrow B) \rightarrow B) \rightarrow ((B \rightarrow A) \rightarrow A) \\
(A4): ( \neg A \rightarrow \neg B) \rightarrow (B \rightarrow A) \\
(A5): ((A \rightarrow B) \rightarrow (B \rightarrow A)) \rightarrow (B \rightarrow A)
\end{align}

\begin{verbatim}
(CD) Cpq : {p->q, p} -> q
(A1) CpCqp
(A2) CCpqCCqrCpr
(A3) CCCpqqCCqpp
(A4) CCNpNqCqp
(A5) CCCpqCqpCqp
\end{verbatim}

It is said that Wajsberg gave proof of this conjecture in the 1930s. But it has been lost (not published).
Afterward, in the 1950s, Rose and Rosser \cite{rosser} provided the proofs.
Also, the completeness of the proofs concerning more general classes of the algebraic structure was presented in \cite{chang}.

\section{Methodology}

\subsection{Hyper-Resolution}
Hyper-Resolution takes a non-positive clause called the nucleus and simultaneously
infers each of its negative literals. Those negative literals are called satellites.

The general scheme is:
\begin{align*}
& {K_{1,l},.,K{1,n}} \\
& ...\\
& {K{m,l},.,K{m,n}}\\
& \underline{ \{\lnot L_1, . , \lnot L_{m+1}, L_{l}\} \exists \sigma.\sigma} \\
& \underline{ = mgu( [ | K_1 |, . , | K_{m,1} |], [|L_1|, . , |L_m|]) } \\
& \{K_{1,2},.,K_{1,n},K_{m,2},.K_{m,n},L_{m+1},.,L_l \} \sigma
\end{align*}

$K_i$ denotes the clauses in the list above, and $L_i$ denotes the literal.
Hyper-Resolution is applied to a set of m unit clauses {K1} ... {Km} and a single nucleus {L1, ..., Lm+1} consisting of m + 1 literals.

\subsection{Condensed detachment}

Condensed detachment \cite{wos} enables us to combine substitutions and modus ponens into a single rule.

\begin{verbatim}
i(s,t) : major premise
r : minor premise
\end{verbatim}

yields $t_\sigma$, where $\sigma$ is a most general unifier for terms r and s.


For the expression of condensed detachment in the language of OTTER, we apply hyper-resolution.
The single rule of condensed detatchment is a particluar instance of a nucleus of the form:

\begin{verbatim}
list(usable).
-P(i(x,y)) | -P(x) | P(y).
end_of_list.
\end{verbatim}

Here the predicate P (provable) is a unary predicate symbol.
If we apply condensed detachment to equation (1), we obtain clause 61 as follows.

\begin{verbatim}
1 [] -P(i(x,y))| -P(x)|P(y).
2 [] P(i(x,i(y,x))).
61 [hyper,1,2,2] P(i(x,i(y,i(z,y)))).
\end{verbatim}

\subsection{Hints: a proof sketch}

Hins in OTTER can be described as a proof sketch.
The proof sketch for yielding A5 from A1-A4 in $\L_{{\aleph}_{0}}$ can be thought of as a set of conditions which is lemmas to prove
sufficient for the proof in $\L_{{\aleph}_{0}}$.
Ideally, a proof sketch composed of a hints list is already proof. Then the conditions are also met.

The hints strategy is closely related to the weighting strategy. Both strategies provide the desired control over the reasoning .process.

Weighting focus on each clause for assigning a weight by upper-supplied weight templates.
Templates provide a recursive mapping of weight from term and atoms corresponding the values of weight as follows:

\begin{verbatim}
weight_list(pick_given).
weight(P(i(x,i(y,x))), 2).
weight(P(i(i(x,y),i(i(y,z),i(x,z)))), 2).
end_of_list.
\end{verbatim}

In contrast, the hints strategy copes with the identification of key clauses in the reasoning process
instead of the general calculation of weights.

\begin{verbatim}
list(hints).
P(i(i(x,y),i(i(z,x),i(z,y))))
end_of_list.
\end{verbatim}

The hint strategy enhances the weighting strategy in two points.

\begin{enumerate}
\item Hints strategy focuses on entire facts (clauses) rather than terms and subterms.
\item Hints strategy uses subsumption for determining the value of a generated clause.
\end{enumerate}

The hints strategy provides both semantic and logical components for the evaluation of a generated clause
by leveraging subsumption.

\section{Experiment}

In experiment, we use workstation with Intel(R) Xeon(R) CPU E5-2620 v4 (2.10GHz) and 251G RAM.
We set the list of set of support corresponding to the equations (1)-(4) as follows:

\begin{verbatim}
list(sos).
P(i(x,i(y,x))).
P(i(i(x,y),i(i(y,z),i(x,z)))).
P(i(i(i(x,y),y),i(i(y,x),x))).
P(i(i(n(x),n(y)),i(y,x))).
end_of_list.
\end{verbatim}

We set the passive list to search unit conflict to yield the proof from (1)-(4) to (5).

\begin{verbatim}
list(passive).
-P(i(i(i(a,b),i(b,a)),i(b,a))).
end_of_list.
\end{verbatim}

Figure 1 shows the average size of the set of support.
In the experiment, we repeated the random extraction of hints 150 times.
Average is calculated for each size of the hints list.
Also, Figure 2 shows the average number of generated clauses.

\begin{align}
average \, size = \frac { \sum_{n=1}^{150} Xn}{150}
\end{align}

The X-axis of Figure 1 and Figure 2 is the size of the hints list, ranging from 30 to 60.
The size of set support shown in Figure 1 is relatively stable compared with Figure 2.
The number of generated clauses is drastically increased between hints*40 to hints*45.

\begin{figure}[htbp]
\centerline{\includegraphics[scale=0.45]{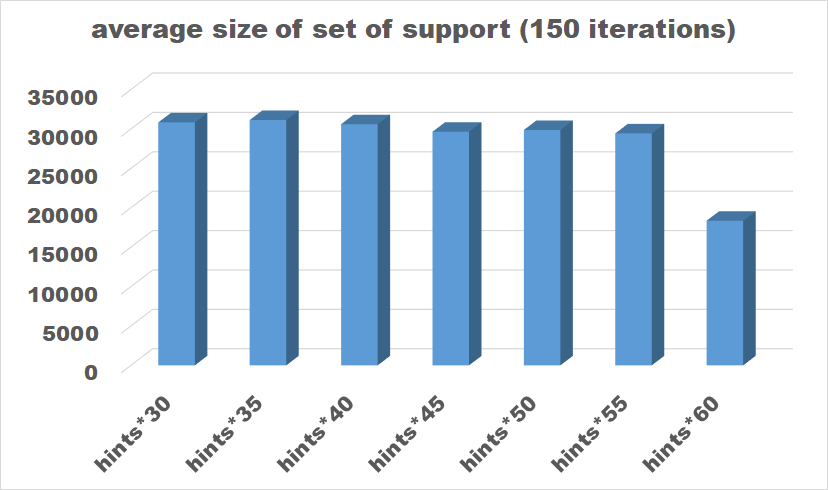}}
\caption{average size of set of support.}
\label{fig}
\end{figure}

\begin{figure}[htbp]
\centerline{\includegraphics[scale=0.45]{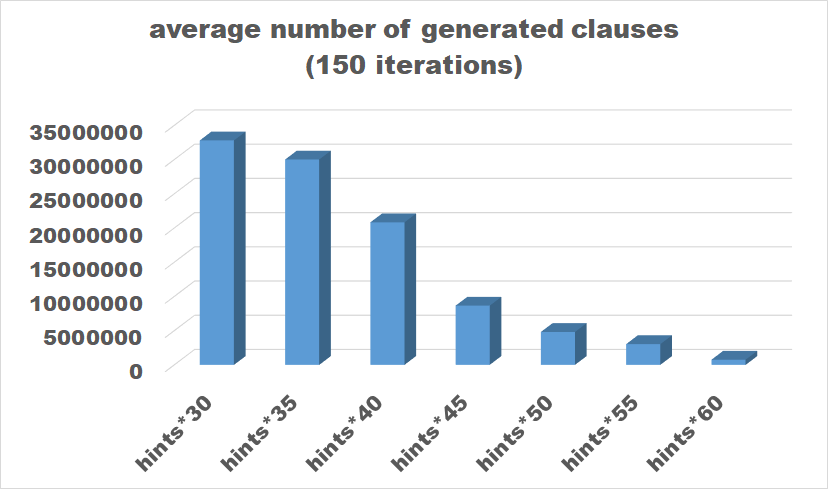}}
\caption{average number of generated clauses.}
\label{fig}
\end{figure}

Figure 3 depicts the size of the set of support for every 150 times of iterations.
We sort the X-axis by the size of the set of support.
Figure 4 depicts the number of generated clauses for every 150 times of iterations.
We sort the X-axis by the number of generated clauses.
In Figure 3, some change points are observed in areas A (x=95-100) and B (x=126-129)
Curiously, there is no difference between hints*30 and hints*45 from x=1 to 95.
In Figure 4, we observe two change points, A (x=84-87) and B (x=119-120).
During these periods, the number of generated clauses with hints*30 and hints*45 are drastically increased.

\begin{figure}[htbp]
\centerline{\includegraphics[scale=0.5]{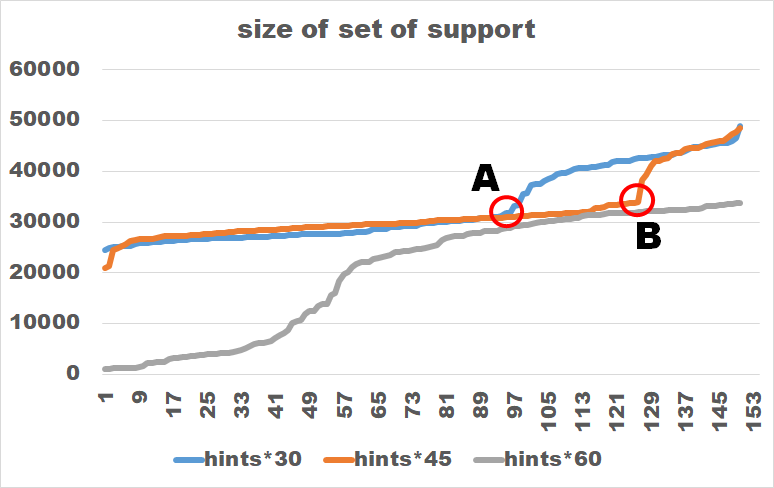}}
\caption{size of set support.}
\label{fig}
\end{figure}

\begin{figure}[htbp]
\centerline{\includegraphics[scale=0.5]{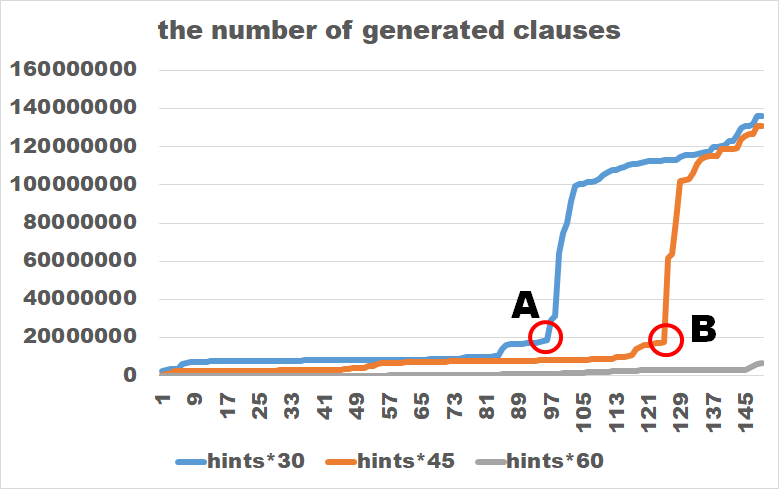}}
\caption{the number of generated clauses.}
\label{fig}
\end{figure}

\section*{References}
We use the automated reasoning program OTTER \cite{otter}.
Veroff present the effectiveness of hints strategy in automated reasoning \cite{b5}.
The axiomatics of L0 is presented in \cite{b6}.
The use of condensed detachment is discussed in \cite{kalman}.

\section{Conclusion}
In this paper, we present an experiment of our randomized hints strategy of automated reasoning
for yielding Axiom 5 on \L ukasiewicz axiom $\L_{{\aleph}_{0}}$.
We have successfully found the most useful 30 hints among 900 hints lists.
Also, we have observed some curious non-linear increases in the size of set of support and the number of generated clauses in deducing Axiom (5) by comparing the number of hints from 30 to 60.

\onecolumn

\appendix
\begin{table}[htb]
\begin{center}
\caption{The most effective 30 hints in proof in \L ukasiewicz axioms $\L_{{\aleph}_{0}}$ .}
\begin{tabular}{|l|l|} \hline
clauses given & 4403 \\ \hline
clauses generated & 256223 \\ \hline
hyper res generated & 256223 \\ \hline
demod \& eval rewrites & 771375 \\ \hline
clauses forward subsumed & 1053187 \\ \hline
(subsumed by sos) & 94343 \\ \hline
clauses kept & 30781 \\ \hline
usable size & 4404 \\ \hline
sos size & 26382 \\ \hline
kbytes malloced & 38085 \\ \hline
\end{tabular}
\end{center}
\end{table}

\begin{verbatim}
list(hints).
P(i(i(i(x,y),z),i(i(n(y),n(x)),z))).
P(i(i(x,y),i(n(y),i(x,z)))).
P(i(i(x,y),i(i(i(y,x),x),y))).
P(i(i(i(x,n(i(x,y))),n(i(y,x))),y)).
P(i(i(i(i(x,y),y),n(i(x,y))),n(y))).
P(i(i(x,i(i(i(y,z),z),n(i(y,z)))),i(x,n(z)))).
P(i(i(n(x),y),i(n(y),x))).
P(i(i(x,i(n(y),n(z))),i(x,i(z,y)))).
P(i(i(x,i(i(i(y,z),z),n(i(z,y)))),i(x,n(y)))).
P(i(i(x,i(n(n(y)),z)),i(x,i(y,z)))).
P(i(i(n(x),n(i(i(y,z),i(i(z,w),i(y,w))))),x)).
P(i(x,i(i(i(y,x),i(x,y)),y))).
P(i(i(i(n(x),i(y,n(i(y,x)))),z),z)).
P(i(i(x,i(i(y,z),w)),i(x,i(z,w)))).
P(i(x,i(i(y,z),i(i(x,y),z)))).
P(i(i(x,i(i(y,n(i(y,z))),n(i(i(y,n(i(y,z))),n(z))))),i(x,z))).
P(i(i(x,y),i(n(y),n(x)))).
P(i(i(n(x),n(i(y,i(z,y)))),x)).
P(i(i(i(i(x,y),i(i(z,x),y)),w),i(z,w))).
P(i(i(i(x,n(i(x,y))),n(i(i(x,n(i(x,y))),n(y)))),y)).
P(i(i(i(x,n(i(x,y))),n(i(i(x,n(i(x,y))),n(y)))),n(n(y)))).
P(i(i(i(n(x),n(y)),n(y)),i(i(x,y),n(x)))).
P(i(i(x,i(y,z)),i(x,i(i(z,w),i(y,w))))).
P(i(i(i(x,y),z),i(n(x),z))).
P(i(i(i(n(n(x)),y),z),i(i(x,y),z))).
P(i(i(i(x,y),i(y,x)),i(y,x))).
P(i(i(x,i(y,z)),i(y,i(x,z)))).
P(i(i(x,n(i(y,z))),i(x,n(i(i(z,w),n(y)))))).
P(i(i(i(i(x,y),y),z),i(i(i(y,x),x),z))).
P(i(n(n(x)),x)).
end_of_list.
\end{verbatim}

\end{document}